\documentclass[12pt]{JHEP3}
\usepackage{amsfonts}
\usepackage{amscd}
\usepackage{amssymb}
\usepackage{amsmath}
\usepackage{epsfig}
\usepackage{latexsym}
\usepackage{mathrsfs}

\numberwithin{equation}{section}
\def \be  {\begin{equation}}
\def \ee  {\end{equation}}
\def \ba  {\begin{eqnarray}}
\def \ea  {\end{eqnarray}}

\newcommand\cA{\mathcal{A}}

\newcommand\cD{\mathcal{D}}
\newcommand\cF{\mathcal{F}}

\newcommand\cN{\mathcal{N}}
\newcommand\cO{\mathcal{O}}

\newcommand\bC{\mathbb{C }}
\newcommand\bP{\mathbb{P}}

\newcommand\bR{\mathbb{R}}

\newcommand\CP {\mathbb{CP}}

\newcommand\rD{\mathrm{D}}
\newcommand\rP{\mathrm{P}}
\newcommand\rU{\mathrm{U}}
\newcommand\rW{\mathrm{W}}

\newcommand\rd{\mathrm{d}}
\newcommand\e{\mathrm{e}}
\newcommand\im{\mathrm{i}}

\newcommand\la{\langle}
\newcommand\ra{\rangle}
\newcommand\del{\partial}
\newcommand\delbar{\bar{\partial}}

\newcommand\MHVbar{\overline{\mbox{MHV}}}

\title{Holomorphic Linking, Loop Equations and Scattering Amplitudes in Twistor Space}

\author{Mathew Bullimore\\
	Rudolf Peierls Institute for Theoretical Physics,\\
	1 Keble Road, Oxford, OX1 3NP,
	United~Kingdom}

\author{David Skinner\\
	Perimeter Institute for Theoretical Physics,\\ 
	31~Caroline~St., Waterloo, ON, N2L 2Y5, 
	Canada}

\abstract{
	We study a complex analogue of a Wilson Loop, defined over a complex curve, in non-Abelian holomorphic Chern-Simons theory. We obtain a version of the Makeenko-Migdal loop equation describing how the expectation value of these Wilson Loops varies as one moves around in a holomorphic family of curves.  We use this to prove (at the level of the integrand) the duality between the twistor Wilson Loop and the all-loop planar S-matrix of $\cN=4$ super Yang-Mills by showing that, for a particular family of curves corresponding to piecewise null polygons in space-time, the loop equation reduce to the all-loop extension of the BCFW recursion relations. The scattering amplitude may be interpreted in terms of holomorphic linking of the curve in twistor space, while the BCFW relations themselves are revealed as a  holomorphic analogue of skein relations.
}

\begin{document}


\section{Introduction}
\label{sec:intro}

Scattering processes are usually defined for external states that each have some definite momentum on the mass-shell. 
Translational invariance implies that such an amplitude is a distribution with support only when
\be
	\sum_{i=1}^n p_i = 0\, .
\label{momcons}
\ee
In planar gauge theories the external states are cyclically ordered, so we can solve this constraint by introducing `region' or `affine' momenta $x_i$ via
\be
	x_i-x_{i+1}=p_i\, .
\label{dualxdef}
\ee
The mass-shell conditions $p_i^2=0$ constrain adjacent $x_i\,$s to be null separated. In this way, the momenta and colour-ordering of the scattering amplitude are encoded in a piecewise null polygon, as shown on the left of figure~\ref{fig:incidence}.

In planar $\cN=4$ super Yang-Mills, it turns out that scattering amplitudes take their simplest form when expressed in (momentum) twistor space.  (See {\it e.g.}~\cite{Penrose:1986ca,HuggettTod,WardWells} for introductions to twistor theory and~\cite{Hodges:2009hk,Mason:2009qx} for details specific to momentum twistors.) In part, this is because (momentum) twistors make manifest the (dual) superconformal symmetry~\cite{Drummond:2008vq} of the tree amplitudes and loop integrand~\cite{Arkani-Hamed:2010kv}, whose existence reflects the Yangian symmetry of the planar theory in the amplitude sector. (See also~\cite{Drummond:2010zv} for very interesting hints of the role of the Yangian in the loop amplitudes themselves.) However, a perhaps deeper reason that these amplitudes belong in twistor space is because {\it their twistor data is unconstrained}. This is because points in space-time correspond to complex lines (linearly embedded Riemann spheres) in twistor space, while two such points $x_1$ and $x_2$ are null separated if and only if their corresponding twistor lines X$_1$ and X$_2$ intersect. Thus, to specify an $n$-sided piecewise null polygon in space-time, we can pick $n$ {\it arbitrary} twistors and sequentially join them up with (complex\footnote{As usual, we draw real lines in three-space to depict complex lines in the complex three-dimensional space.}) lines.   Adjacent lines intersect by construction, so that as well as momentum conservation, the mass-shell condition is automatic.  Thus the scattering data is encoded in a picture that is usually drawn as on the right of figure~\ref{fig:incidence}, where the twistors $z_i$ are chosen freely.

\FIGURE[t]{
	\includegraphics[height=45mm]{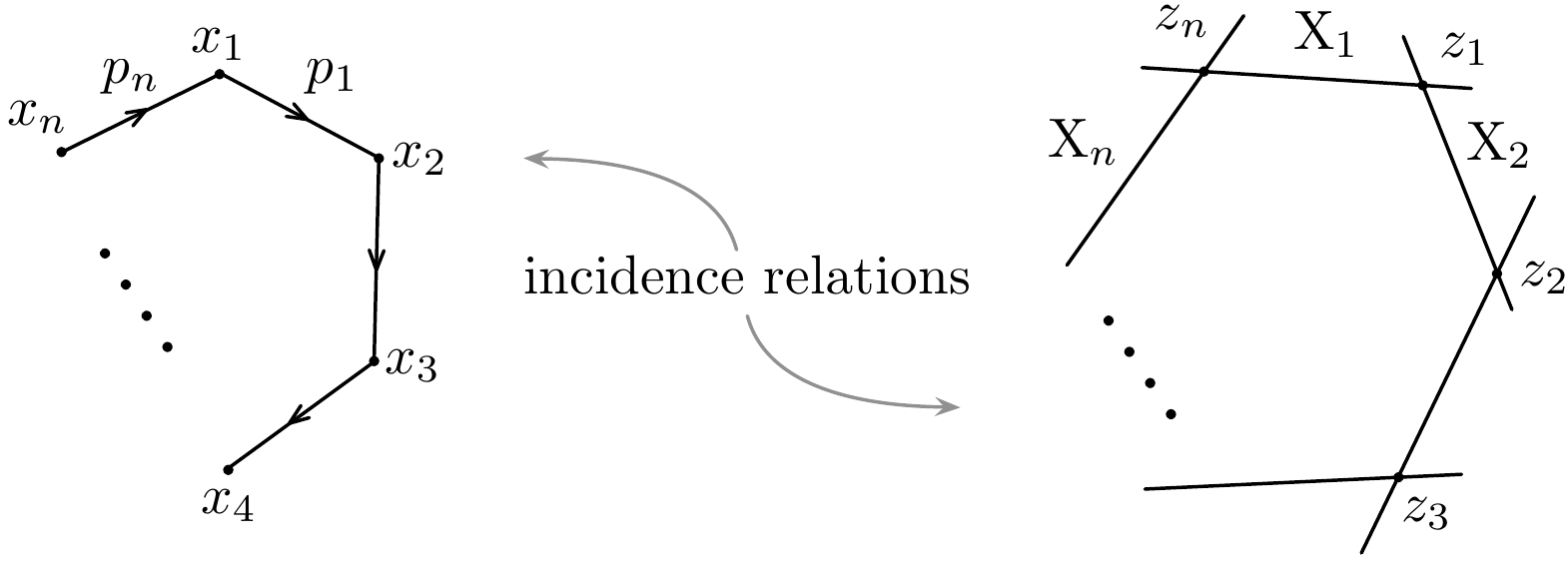}
	\caption{A depiction of (a planar projection of) a nodal curve in twistor space corresponding to a piecewise null 
	polygon in space-time. Via the incidence relations, the vertex $x_i$ corresponds to the line ${\rm X}_i=(z_{i-1},z_i)$ 
	while the node $z_i$ corresponds to the null ray through $x_i$ and $x_{i+1}$.}
	\label{fig:incidence}
}

\medskip

\begin{figure}[t]
\begin{center}
	\includegraphics[width=65mm]{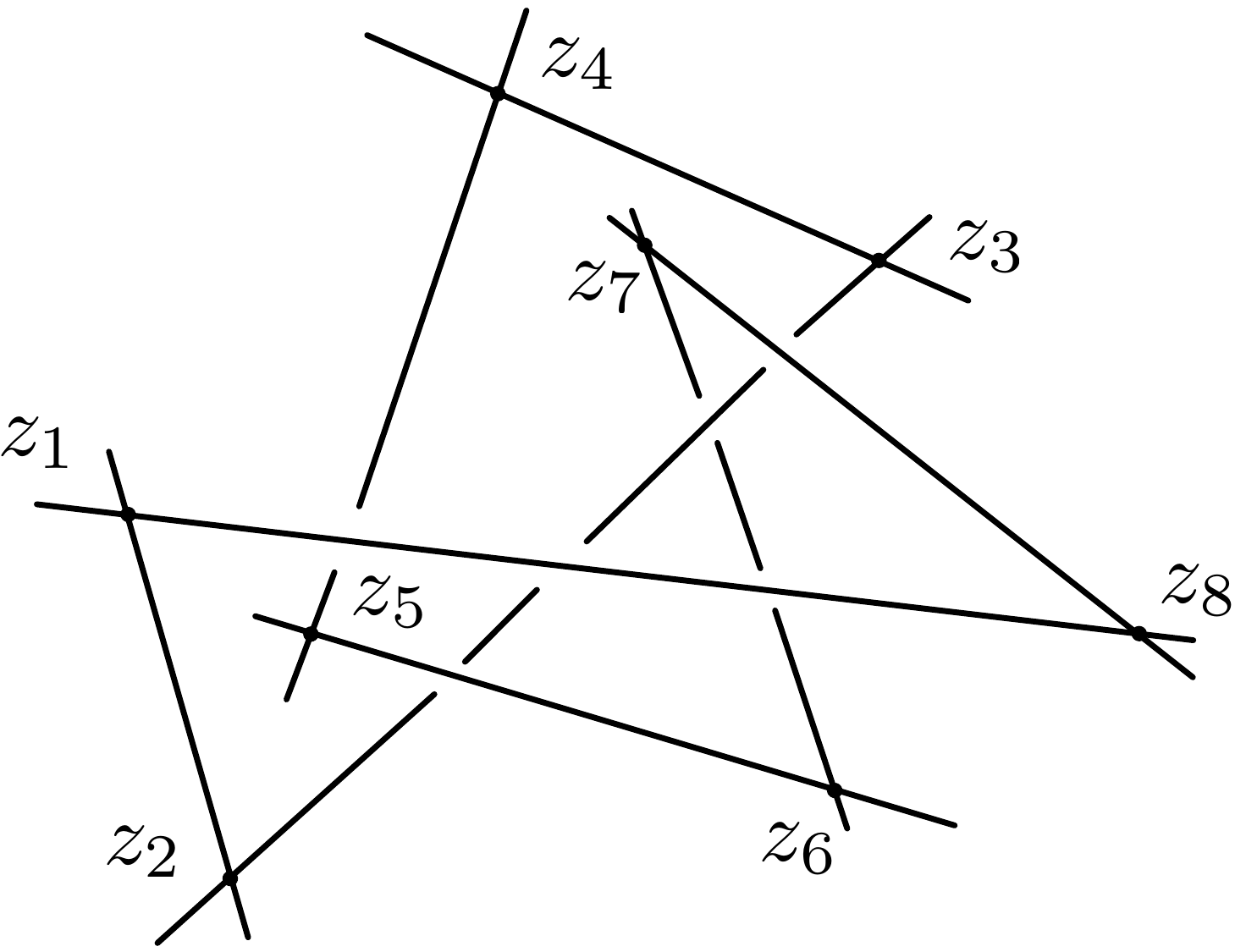}
	\caption{Depending on the location of the vertices, the twistor lines may form a polygonal knot. The plane 
	projection of the nodal curve can thus be far more involved than figure 1 suggests.}
	\label{fig:knot}
\end{center}
\end{figure}

The key observation that motivates this paper is that although figure~\ref{fig:incidence} always provides an accurate representation of the twistor curve as an abstract Riemann surface, the holomorphic embedding of this Riemann surface in a complex three-fold can be far more complicated than this picture indicates. Depending on the choice of vertices $z_i$, {\it i.e.} depending on the external kinematics, the embedded twistor curve can also appear as depicted in figure~\ref{fig:knot}: that is, {\it it appears as a polygonal knot}\footnote{Planarity of the colour-ordered amplitude has nothing to do with the embedding of the curve.}.

The new picture hints at an intriguing connection between scattering amplitudes and a complexification of knot theory known as {\it holomorphic linking}. To understand this notion, observe by a simple dimension count that while one can always use smooth deformations to untangle real two-surfaces in a real six-manifold, it is not always possible to untangle complex curves in a complex three-fold if one is only allowed to vary the curve holomorphically. (This restriction provides a second\footnote{The standard reason is that it gives a better picture of the structure of the nodes; pictures of 2-surfaces incorrectly suggest that the component curves are tangent at the nodes.} justification for the use of the half-dimensional representation in these figures.)   That holomorphic linking in twistor space should encode interesting information about gauge theories in space-time was originally envisaged by Atiyah~\cite{Atiyah:1981ey} and Penrose~\cite{PenroseTN27}.  Outside the twistor context, holomorphic linking (and its relation to Abelian holomorphic Chern-Simons theory) has been studied by {\it e.g.} Khesin and Rosly~\cite{Khesin:2000ng} and by Frenkel and Todorov~\cite{Frenkel:2005qk}.

\medskip

It is the aim of this paper to elucidate and further investigate this link. Considerable evidence supporting the existence of such a relation comes from a conjecture given by Mason and one of the present authors in~\cite{Mason:2010yk}, building on previous work by Alday and Maldacena~\cite{Alday:2007hr} using the AdS/CFT correspondence and by Drummond, Korchemsky and Sokatchev~\cite{Drummond:2007aua}, and by Brandhuber, Heslop and Travaglini~\cite{Brandhuber:2007yx} in space-time. The conjecture states, firstly, that the complete {\it classical} S-matrix of $\cN=4$ super Yang-Mills is equivalent to the expectation value in the QFT defined by the holomorphic Chern-Simons functional~\cite{Witten:1992fb,Witten:2003nn}
\be
	S_{\rm hCS}[\cA] =\frac{1}{g_s}\int_{\CP^{3|4}} \Omega\wedge
	{\rm tr}\left(\frac{1}{2}\cA\wedge\delbar \cA + \frac{1}{3}\cA\wedge\cA\wedge\cA\right)
\label{hCSintro}
\ee
of a certain operator defined over the complex curve $C$ of figure~\ref{fig:knot}. As we show in section~\ref{sec:WL}, this operator is  a natural complexification of the standard Wilson Loop operator, given by the explicit formula
\be
	\rW_{\!R}[C] = {\rm Tr}_R\,{\rm P}\exp \left(-\int_C\omega\wedge\cA\right)
\ee
in terms of the connection (0,1)-form $\cA$, where the precise definition of the meromorphic form $\omega$ and the meaning of the `path-ordering' symbol P is explained below. The expectation value $\big\la\rW_{\!R}[C]\big\ra_{\rm hCS}$ is thus a natural complexification of the expectation value 
\be
	\big\la W_{\!R}[\gamma]\big\ra 
	= \int	\cD A\ \e^{ \im S_{\rm CS}[A]}\ {\rm Tr}_R\,{\rm P}\exp\left(-\oint_\gamma A\right)
\label{real<W>}
\ee 
of a real Wilson loop in real Chern-Simons theory, that famously computes knot invariants~\cite{Witten:1988hf} such as the HOMFLY polynomial (for gauge group U$(N)$ and $R$ the fundamental representation).

\medskip

Extrema of~\eqref{hCSintro} correspond to holomorphic bundles on\footnote{The twistor superspace $\CP^{3|4}$ may be defined as the total space of $\left(\Pi\cO_{\bP^3}(1)\right)^{\oplus 4}$, where $\cO_{\bP^3}(1)$ is the dual of the tautological line bundle on $\CP^3$ and $\Pi$ reverses the Grassmann parity of the fibres. It is a Calabi-Yau supermanifold and we have written $\Omega$ for the canonical holomorphic section of its Berezinian.} $\CP^{3|4}$ that (under mild global conditions) in turn yield anti self-dual (or 1/2 BPS) solutions of the $\cN=4$ super Yang-Mills equations on space-time via the Penrose-Ward correspondence. Thus, holomorphic Chern-Simons theory corresponds only to the anti self-dual sector of $\cN=4$ SYM. However, it is possible to add a term to~\eqref{hCS} so that the new action corresponds to the complete (perturbative) $\cN=4$ theory~\cite{Boels:2006ir}. The second part of the conjecture of~\cite{Mason:2010yk} is that, when weighted by the exponential of this new action, $\big\la\rW[C]\big\ra$ now yields the complete planar S-matrix of $\cN=4$ super Yang-Mills,  to all orders in the 't Hooft coupling and for arbitrary external helicities.

These conjectures were supported in~\cite{Mason:2010yk} by perturbative calculations showing that, at least for small $k$ and $\ell$, the axial gauge Feynman diagrams of $\big\la\rW[C]\big\ra$ are in one-to-one correspondence with MHV diagrams for the $\ell$-loop N$^k$MHV scattering amplitudes. The momentum twistor formulation of the MHV diagram formalism had previously been given in~\cite{Bullimore:2010pj} and it has since been proved~\cite{Bullimore:2010dz,He:2010ju} that this formalism correctly reproduces the all-loop recursion relation for the integrand of planar $\cN=4$~\cite{Arkani-Hamed:2010kv,Boels:2010nw}. 

\medskip

In this paper, after reviewing the construction of the Wilson Loop operator for a complex curve in section~\ref{sec:WL}, we study a complex analogue of the Makeenko-Migdal loop equations~\cite{Makeenko:1979pb}, for both the holomorphic Chern-Simons and $\cN=4$ SYM theories. At least in principle, these equations give a non-perturbative description of the behaviour of $\big\la\rW[C]\big\ra$ as $C$ varies holomorphically.  

Now, the loop equations for the expectation value~\eqref{real<W>} of a fundamental Wilson Loop in real U($N$) Chern-Simons theory were shown by Cotta-Ramusino, Guadagnini, Martellini and Mintchev~\cite{CottaRamusino:1989rf} to yield the skein relations for the HOMFLY polynomial -- in other words, recursion relations allowing one to reconstruct the knot invariants. Quite remarkably, in section~\ref{sec:amplitudes} we find that (in the planar limit) the loop equations for $\big\la\rW[C]\big\ra$ may be reduced to the BCFW relation~\cite{Britto:2005fq} that recursively constructs the classical S-matrix. The classical BCFW recursion relations thus emerge as a holomorphic analogue of the skein relations for a knot polynomial! Replacing the pure Chern-Simons theory by the full $\cN=4$ theory, the loop equations involve a new term and reduce instead to the recursion relation of Arkani-Hamed, Bourjaily, Cachazo, Caron-Huot and Trnka~\cite{Arkani-Hamed:2010kv} for the all-loop integrand. (See also~\cite{Boels:2010nw} for tentative extensions of this relation beyond $\cN=4$ SYM.)  These results prove the conjectures of~\cite{Mason:2010yk}, at least at the level of the integrand. 

Working in space-time, Caron-Huot demonstrated in~\cite{Caron-Huot:2010ek} that the expectation value in $\cN=4$ SYM of a certain extension of a regular Wilson Loop around the null polygon on the left of figure~\ref{fig:incidence}, together with extra operator insertions at the vertices, also obeys the all-loop BCFW recursion relation. This complete definition of this Wilson Loop operator and the vertex insertions was not given in~\cite{Caron-Huot:2010ek}, but it is expected to be fixed by supersymmetry.  Our proof that the twistor Wilson Loop (with no extra insertions) obeys the same relation helps confirm that these two approaches agree.

There are at least two ways to view these results. On the one hand, accepting the duality between scattering amplitudes and Wilson Loops, one may see the loop equations as providing an independent derivation of the all-loop BCFW recursion relations. However, we prefer to interpret this work as a field theoretic explanation for {\it why} the duality holds in the first place: the behaviour of a scattering amplitude near a factorisation channel is exactly equal to the behaviour of the twistor Wilson Loop near a self-intersection. Away from these singularities, both the scattering amplitudes and Wilson Loops are determined by analytic continuation of their data. For scattering amplitudes this is just the analyticity of the S-matrix as a function of on-shell momenta, encoding the familiar notions of crossing symmetry and space-time causality, while for Wilson Loops, analyticity in the twistors $z_i$ expresses the fact that $\big\la\rW[C]\big\ra$ generates analytic invariants associated to the holomorphic link.

Finally, although it is often said that BCFW recursion relations {\it for scattering amplitudes} have little to do with Lagrangians, path integrals or gauge symmetry, our work makes it clear that these standard tools of QFT are indeed intimately connected to BCFW recursion {\it from the point of view of Wilson Loops}. As always, the Wilson Loop is just the trace of the holonomy of a connection around a curve, while the loop equations themselves are derived from little more than the Schwinger-Dyson equations of the appropriate path integral. We view the connection to Lagrangians and path integrals as a virtue, because it suggests that BCFW-type techniques can be applied to observables beyond scattering amplitudes. Certainly, the loop equations we derive are valid for far more general curves than needed for scattering amplitudes, and for more general deformations than BCFW.


\section{Non-Abelian Holomorphic Wilson Loops}
\label{sec:WL}

\begin{figure}
\begin{center}
	\includegraphics[width=65mm]{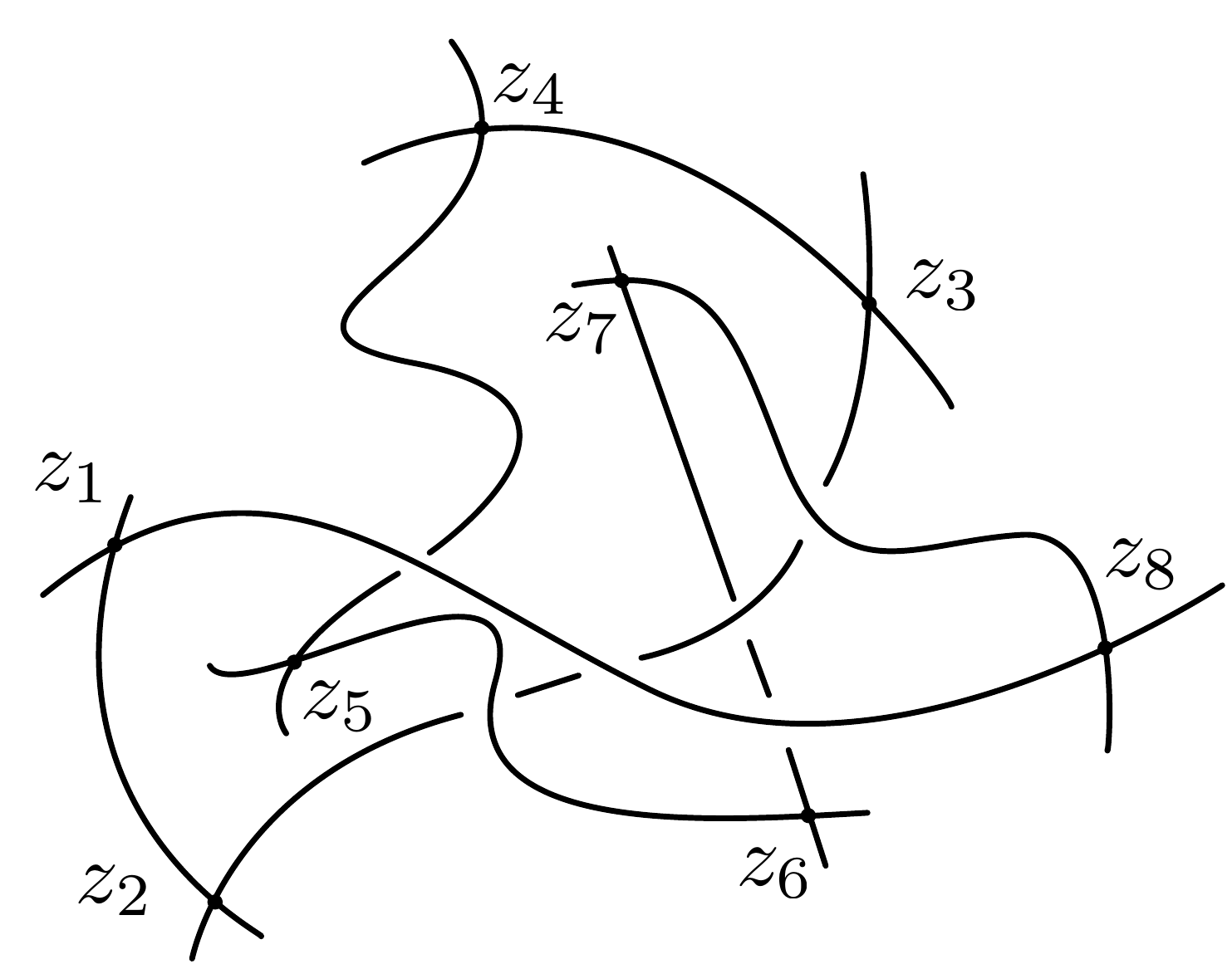}
	\caption{Going beyond the application to scattering amplitudes, we may allow the component curves to 
	have arbitrary degrees $\geq1$.}
	\label{fig:curvyknot}
\end{center}
\end{figure}

In this section we review the construction of~\cite{Mason:2010yk} for a complex Wilson Loop defined over a nodal curve $C$, each of whose components is rational. That is, $C=Z(\Sigma)$ where $Z$ is a holomorphic map to $\CP^{3|4}$ and the source curve
\be
	\Sigma = \Sigma_n\cup\cdots\cup\Sigma_2\cup\Sigma_1
\ee
with each component $\Sigma_i$ rational and $\Sigma_{i+1}\cap\Sigma_i=\{\rm pt\}$ ($i$ being counted mod $n$).  For scattering amplitudes, we only need to consider the case that $Z$ restricts to a linear map on each irreducible component, so that $C$ generically appears as in figure~\ref{fig:knot}. However, it costs little to extend this to the case where each component is mapped with arbitrary degree $d_i\geq1$ (shown in figure~\ref{fig:curvyknot}) and we allow this more general possibility. These curves may also be thought of as elliptic curves that have been pinched $n$ times around a given homology cycle. They are certainly rather exotic from the algebro-geometric point of view\footnote{However, it is curious that in the real category, one often considers `tame' knots -- {\it i.e.}, knots that are equivalent to an embedded polygon.}, and it would clearly be interesting consider the extension to non-degenerate elliptic curves.

\subsection{Holomorphic frames}
\label{sec:h}

Any $C^\infty$ bundle $E\to \CP^{3|4}$ is inevitably holomorphic on restriction to a arbitrary complex curve $C$, because $\bar D^2|_C=0$ trivially for dimensional reasons. Suppose first that $C$ is a single, irreducible rational curve. Provided $E|_C$ is holomorphically trivial\footnote{By a theorem of Birkhoff and Grothendieck, $E|_C = \bigoplus\cO(a_i)$ with $\sum a_i=0$ if $E|_C$ is topologically trivial. Triviality as a holomorphic bundle means that, in addition, $a_i=0$ for each $i$ and (perhaps surprisingly) is in fact generic.}, we can find a gauge transformation $h$, varying smoothly over $C$ and such that
\be
	h^{-1}\circ\left.(\delbar+\cA)\right|_C\circ h = \left.\delbar\right|_C\, .
\label{hgauge}
\ee
Under this gauge transformation, covariantly holomorphic objects become simply holomorphic on $C$, so $h$ is said to define a {\it holomorphic frame} for $E|_C$. It follows from~\eqref{hgauge} that $h$ itself obeys
\be
	\left.(\delbar+\cA)\right|_C h = 0
\label{framesol}
\ee
and so is defined up to a gauge transform $h \to hh'$, where $h'$ must be globally holomorphic on $C$.

For two points $z_0,z\in\CP^{3|4}$ that lie on $C$, we now define 
\be
	\rU(z,z_0) \equiv h(z)h^{-1}(z_0) 
\label{Udef}
\ee
to be the particular solution of~\eqref{framesol} obeying the initial condition that the gauge transformation at $z_0\in C$ is simply the identity. The holomorphic frame $\rU(z,z_0)$ defines a map
\be
	\rU(z,z_0): E|_{z_0} \to E|_z
\ee
between the fibres of $E$ at $z_0$ and  $z\in \CP^{3|4}$, that depends on the choice of curve $C$ (not unique if ${\rm deg}(C)>1$). It is the natural complex analogue of the parallel propagator $U(x,x_0)$ between two points $x$ and $x_0$ along a real curve $\gamma$. For example, it follows immediately from the definition~\eqref{Udef} that 
\be
	\rU(z,z')\rU(z',z_0) = \rU(z,z_0)\,,
\label{Uconcatenate}
\ee
so that $\rU(z,z_0)$ concatenates, just as for the parallel propagator along a real curve. In particular, $\rU(z,z_0)\rU(z_0,z) = \rU(z,z) = 1_{R}$, implying that
\be
	\rU(z_0,z) = {\rU(z,z_0)}^{-1}\, .
\label{Uinverse}
\ee
Likewise, if we change our choice of connection (0,1)-form by a smooth gauge transformation
\be
	\cA\to\cA_g = g\,\delbar g^{-1} + g\cA g^{-1}\,,
\ee
then since $h^{-1}\circ\left.(\delbar+\cA)\right|_C\circ h = (gh)^{-1}\circ\left.(\delbar+\cA_g)\right|_C\circ(gh)$ we find that
\be
	\rU(z,z_0) \to g(z)\rU(z,z_0)g^{-1}(z_0)
\label{Ugauge}
\ee
again as for a parallel propagator in the real category.

In the Abelian case with $E$ a line bundle, setting $h=\e^{-\phi}$ for some smooth function $\phi$ on $C$, the holomorphic frame equation~\eqref{framesol} becomes $\delbar\phi = \cA|_C$. This equation always has a solution when $C$ is rational, given by
\be
	\phi(z) = \int_C\omega\wedge \cA\,,
\label{phidef}
\ee
where $\cA$ is restricted to the curve and where $\omega$ is a meromorphic 1-form on $C$ whose only singularities are simple poles at $z_0$ and $z$, with residues $\pm1$, respectively. Explicitly, if $C=Z(\Sigma)$ with $Z(\sigma)=z\in\CP^{3|4}$ and $Z(\sigma_0)=z_0$, then the pullback of $\omega$ to $\Sigma$ is
\be
	\omega(\sigma') = \frac{(\sigma-\sigma_0)}{(\sigma-\sigma')(\sigma'-\sigma_0)}\,\frac{\rd \sigma'}{2\pi\im}
\label{merodiff}
\ee
in terms of local coordinates $\sigma\in\Sigma$. This is just the Green's function for the $\delbar$-operator acting on smooth functions on $\Sigma$. The holomorphic frame equation fixes $\phi$ only up to a globally holomorphic function, which by Liouville's theorem must be constant. The constant is fixed by the choice of reference point: $\omega$ vanishes if $z$ and $z_0$ coincide so~\eqref{phidef} has $\phi(z_0)=0$. Thus, in the Abelian case we have the explicit expression
\be
	\rU(z,z_0) = \e^{\phi}= \exp\left(-\int_C\omega\wedge \cA\right)
\label{AbelianU}
\ee
for our holomorphic frame.

In the non-Abelian case, acting on a representation $R$ of the gauge group, $\rU(z,z_0)$ may similarly be given by the (somewhat formal)  expression\footnote{An equivalent formal expression, given in~\cite{Mason:2010yk}, is $\rU(z,z_0) =  \left(\left.1+\delbar^{-1}\cA\right|_{C}\right)^{-1}$, where in the series expansion of this expression, the inverse $\delbar$-operators are taken to act on everything to their right.}
\be
	\rU(z,z_0)={\rP}\exp\left(-\int_C\omega\wedge \cA\right)\,.
\label{Pexp}
\ee
The meaning of the `path ordering' symbol P on our complex curve is that the $i^{\rm th}$ power of the meromorphic differential $\omega$ that appears from expanding the exponential in~\eqref{Pexp} should be taken to have its simple poles located at the reference point $z_0$ and at either the insertion point of the  $(i+1)^{\rm th}$ power of the field $\cA$ (counting according to the colour-ordering), or else at the final evaluation point $z$.  Specifically, 
\be
	\rP\,\exp\left(-\int_C\omega\wedge \cA\right) 
	\equiv 1_R + \,\sum_{m=1}^\infty (-1)^m \int\limits_{(C)^m} 
	\bigwedge_{i=1}^m\  \Big(\omega(z_i)\wedge\cA(z_i)\Big)
\label{Pexp2}
\ee
where $\omega(z_i)$ is a meromorphic differential in $z_i$ with a simple pole at $z_0$ and a simple pole at $z_{i+1}$ (or at $z$ when $i=m$). Explicitly, in terms of intrinsic coordinates,
\be
	\bigwedge_{i=1}^m \omega(\sigma_i) 
	= \frac{(\sigma-\sigma_0)}{(\sigma-\sigma_m)\cdots(\sigma_2-\sigma_1)(\sigma_1-\sigma_0)}\, 
	\frac{\rd \sigma_m}{2\pi\im}\wedge\cdots\wedge\frac{\rd \sigma_1}{2\pi\im}\, .
\ee

Equation~\eqref{Pexp} is thus analogous to the (equally formal) expression
\be
\begin{aligned}
	&\rP\,\exp\left(-\int_{x_0}^x A\right)\\
	 &\qquad = 1_R\ -\ \int_{x_0}^x \rd y^\mu A_\mu(y)
	\ +\  \int_{x_0}^x\!\rd y^\mu A_\mu(y ) \int_{x_0}^y\!\rd {y'}^\nu A_\nu(y')\ - \ \cdots\\
	&\qquad = 1_R + \sum_{m=1}^{\infty}\,(-1)^m\hspace{-0.2cm}
	 \int\limits_{([0,1])^m}\hspace{-0.2cm} \rd s_m \cdots \rd s_1\ 
	A_\mu(s_m)\frac{\rd x^\mu}{\rd s_m}\,\theta(s_m\!-\!s_{m-1})\cdots \theta(s_2-s_1)\,A_\nu(s_1)\frac{\rd x^\nu}{\rd s_1}
\end{aligned}
\ee	
for the parallel propagator of a connection $\rd+A=\rd x^{\mu}(\del_\mu+A_\mu)$ along a real curve $\gamma$, where the step functions obey
\be
	\frac{\del}{\del s}\theta(s-s') =  \delta(s-s')
\ee
and are the Green's function for the exterior differential on the interval $[0,1]$. We also note that the iterated integrals appearing in~\eqref{Pexp} are somewhat similar to iterated integrals that arise in studying parallel transport via the Knizhnik-Zamolodchikov connection over the configuration space of $n$-points on a rational curve. See~\cite{Abe:2009} for a recent discussion of this in a twistor context.


\subsection{The holonomy around a nodal curve}
\label{sec:holonomy}

Having identified the holomorphic frame $\rU(z,z_0)=h(z)h^{-1}(z_0)$ as the natural analogue of a parallel propagator for a complex curve, we can define the holonomy of the connection (0,1)-form around $C$, based at some point $z\in\CP^{3|4}$, to be the ordered product
\be
	{\rm Hol}_{z}[C] \equiv \rU(z,z_n)\,\rU(z_n,z_{n-1})\cdots\rU(z_2,z_1)\,\rU(z_1,z)\,,
\label{Holzdef}
\ee
where $z_i$ is the image of the node $\Sigma_{i+1}\cap\Sigma_i$ and where, without any essential loss of generality,  we have taken the base point to be on $Z(\Sigma_1)$. Reading from the right of~\eqref{Holzdef}, this definition uses the holomorphic frame $\rU(z_1,z)$ on $Z(\Sigma_1)$ to map the fibre $E|_{z}$ to the fibre $E|_{z_1}$ at the node $z_1$. We can identify this fibre with a fibre of $E$ restricted to the next component line $Z(\Sigma_2)$, and then map to the following node using the holomorphic frame $\rU(z_2,z_1)$ on $Z(\Sigma_2)$.  We continue transporting our fibre around the nodal curve in this way until the final factor of $\rU(z,z_1)$ returns us to the starting point. The result is thus an automorphism of the fibre $E|_{z}$, as required.

The holomorphic Wilson Loop is now defined as in the real category as
\be
	\rW_{\!R}[C] \equiv {\rm Tr}_R\, {\rm Hol}_{z}[C]\,,
\label{hWL}
\ee
where to take the trace in a representation $R$ we have used the fact that a holomorphic frame for $E|_{C}$ induces a holomorphic frame for all tensor products of $E|_C$ and its dual, in other words for all choices of representation. Cyclicity of the trace and the concatenation property~\eqref{Uconcatenate} immediately shows that the Wilson Loop is independent of the base point. 

For an Abelian gauge group ($E$ a line bundle) the holomorphic frame~\eqref{AbelianU} gives
\be
	\rW[C] = \exp\left(-\int_C\omega\wedge \cA\right)
\ee
where $\omega$ is a meromorphic form on $C$ whose only singularities are simple poles at each of the nodes, with residues $\pm1$ on each component. Such a differential is nothing but the holomorphic differential $\theta$ on a non-singular elliptic curve in the limit that this curve degenerates to our nodal curve (see {\it e.g.}~\cite{Faybook}). Thus, in the Abelian case we recover the prescription 
\be
	\rW[C] =\exp\left(-\int_C\theta\wedge\cA\right)
\label{AbWL}
\ee
that was proposed by Thomas~\cite{Thomas:1998uj}, by Khesin and Rosly~\cite{Khesin:2000ng} and by Frenkel and Todorov~\cite{Frenkel:2005qk}. For a non-Abelian gauge group we can likewise use~\eqref{Pexp} to formally write
\be
	\rW_{\!R}[C] = {\rm Tr}_R\,\rP\exp\left(-\int_C\omega\wedge \cA\right)
\label{hWLformal}
\ee
as a natural complexification of the familiar expression
\be
	W_{\!R}[\gamma] = {\rm Tr}_{R}\,\rP\exp\left(-\int_\gamma A\right)
\label{realWL}
\ee
for a Wilson Loop of a connection on a real curve $\gamma$. In equation~\eqref{hWLformal}, as in~\eqref{Pexp}, the path ordering symbol means that for each component of $C$, successive powers of $\omega$ in the series expansion of~\eqref{hWLformal} should have one simple pole at one of the nodes and another at the subsequent (with respect to the colour-ordering) insertion point of the field. 


\section{A Holomorphic Family of Curves}
\label{sec:holfamily}

We now investigate the behaviour of our complex Wilson Loop as the curve $C$ varies. We shall prove that, just as the change in the real Wilson Loop~\eqref{realWL} under a smooth deformation of a real curve $\gamma$ is 
\be
	\delta W_{\!R}[\gamma] = -\oint_{\gamma} \rd x^\mu\,\delta x^\nu\,
	{\rm Tr}_R\,  \Big(  F_{\mu\nu}(x)\,{\rm Hol}_{x}[\gamma] \Big)\,,
\label{rWLvary}
\ee
so too the complex Wilson Loop~\eqref{hWL} obeys
\be
	\bar\delta\rW_{\!R}[C] 
	= -\int_C \omega(z)\wedge\rd\bar{z}^{\bar\alpha}\wedge\bar\delta\bar z^{\bar\beta}\ 
	{\rm Tr}_{R}\Big(\cF_{\bar\alpha\bar\beta}(z,\bar z)\,{\rm Hol}_{z}[C]\Big)
\label{hWLvary}
\ee
as the complex curve $C$ moves over a holomorphic family. In this equation,  $\bar\delta$ is the $\delbar$-operator on the parameter space of the family and, once again, $\omega(z)$ is a meromorphic differential on $C$ whose only singularities are simple poles at the nodes with residues $\pm1$.  Like its real analogue, equation~\eqref{hWLvary} expresses the change in the Wilson Loop in terms of the curvature (0,2)-form $\cF=\delbar\cA + \cA\wedge\cA$  through the complex 2-surface swept out by the varying curve. The key point here is that because the curve $C$ varies holomorphically, only the (0,2)-form part of the curvature arises. In particular, if $E$ is a holomorphic bundle on twistor space, then the Wilson Loop is also holomorphic. As before, we prove~\eqref{hWLvary}  in more generality than we actually need for scattering amplitudes, allowing each irreducible component $C_i=Z(\Sigma_i)$ to have degree $d_i\geq1$. 

\medskip

Here is the proof\footnote{It is also possible to show~\eqref{hWLvary} by working directly with the formal expression~\eqref{hWLformal}, provided (as in the real case) due care is taken of the `path ordering'.}. For a fixed partial connection, the holomorphic frame $\rU(z_1,z_0)$ depends on the choice of two points $\{z_0,z_1\}\in\CP^{3|4}$ together with a choice of (irreducible, rational) curve $C$ that joins them. (Note that this curve is not unique if its degree is greater than 1.)  Equivalently, we can think of this data in terms of an abstract Riemann sphere $\Sigma$ together with two marked points $\{\sigma_0,\sigma_1\}\in\Sigma$ and a degree $d$ holomorphic map $Z:\Sigma\to \CP^{3|4}$, such that $Z(\Sigma)=C$ and $Z(\sigma_i)=z_i$. Now suppose we have a holomorphic family of such maps, parametrized by a moduli space $\mathscr{B}$ described, say, by local holomorphic coordinates $t^i$. Writing $\rU(\sigma,\sigma_0;t)$ for the pullback of $\rU(z,z_0)$ to the abstract curve $\Sigma$, for our family of curves the holomorphic frame equation is
\be
	(\delbar+Z^*\cA)\rU(\sigma,\sigma_0;t) = \rd\bar\sigma\left(\frac{\del}{\del\bar\sigma}
	+\cA_{\bar\alpha}\frac{\del\bar z^{\bar\alpha}}{\del\bar\sigma}(\sigma,t)\right)\rU(\sigma,\sigma_0;t)=0
\label{Ufamily}
\ee
where we emphasise that both the holomorphic frame and the connection depend on $t$ -- {\it i.e.}, on the choice of curve.

Consider the integral
\be
	\int_\Sigma \omega_{1,0}(\sigma)\wedge \rU(\sigma_1,\sigma;t)\,(\delbar+Z^*\!\cA)\rU(\sigma,\sigma_0;t)
\label{step0}
\ee
over $\Sigma$, where $\omega_{1,0}(\sigma)$ is our familiar meromorphic differential~\eqref{merodiff} whose only singularities are simple poles at  $\sigma_0$ and $\sigma_1$, with residues $+1$ and $-1$ respectively, and where $\delbar+\cA$ is as  given in~\eqref{Ufamily}.  In principle, the integral depends on these two points and on the map $Z$ -- that is, it depends on the moduli space -- but of course it actually vanishes identically on $\mathscr{B}$ since $\rU(\sigma,\sigma_0)$ obeys~\eqref{Ufamily}.  

Letting $\bar\delta$ be the $\delbar$-operator on $\mathscr{B}$ we have
\be
\begin{aligned}
	0&=\bar\delta\left[\int_{\Sigma} \omega_{1,0}(\sigma)\wedge 
	\rU(\sigma_1,\sigma;t) \left(\delbar+Z^*\!\cA\right) \rU(\sigma,\sigma_0;t)\right]\\
	& = \int_{\Sigma} \omega_{1,0}(\sigma)\wedge \rU(\sigma_1,\sigma;t) \left(\delbar+Z^*\!\cA\right)
	\bar\delta\rU(\sigma,\sigma_0;t)\\
	&\hspace{5cm}
	 -\int_{\Sigma} \omega_{1,0}(\sigma)\,  \rU(\sigma_1,\sigma;t)\wedge
	 \bar\delta\!\left(Z^*\!\cA\right) \rU(\sigma,\sigma_0;t)\\
	 &=-\bar\delta\rU(\sigma_1,\sigma_0;t) 
	 -\int_{\Sigma} \omega_{1,0}(\sigma)\,  \rU(\sigma_1,\sigma;t)\wedge
	 \bar\delta\left(Z^*\!\cA\right)\rU(\sigma,\sigma_0;t)\,,
\end{aligned}
\label{step1}
\ee
where the third line follows from integrating by parts in the first term, then using the poles of $\omega_{1,0}$ and the fact that $\delbar\rU(\sigma_1,\sigma;t) = \rU(\sigma_1,\sigma;t)Z^*\!\cA$ to perform the integral\footnote{In fact, the moduli space $\mathscr{B}$ may be identified with the Kontsevich moduli space $\overline{M}_{0,2}(\CP^{3|4},d)$ of stable, 2-pointed, degree $d$ maps. Because it depends on {\it three} points $\{\sigma_0,\sigma_1,\sigma\}\in\Sigma$ as well as $Z$, the {\it integrand} in~\eqref{step0} really lives on the universal curve $\mathscr{C}\equiv\overline{M}_{0,3}(\CP^{3|4},d)$ and the pullback by the map $Z$ is more properly written as the pullback by the evaluation map on the third marked point (labelled $\sigma$). The integral itself is really the pushdown by the forgetful map  $\pi:\mathscr{C}\to\mathscr{B}$ and in commuting $\bar\delta$ past the integral sign in~\eqref{step1}, we really mean the pullback $\pi^*\bar\delta=\delbar_\mathscr{C}-\delbar_{\mathscr{C}/\mathscr{B}}=\delbar_\mathscr{C}-\delbar_\Sigma$. Then~\eqref{dBofA} is just the statement that the evaluation map is holomorphic and so commutes with $\delbar_{\mathscr{C}}$, with the $\delbar_\Sigma$ term left over.}.

In the final term of~\eqref{step1} we have 
\be
	 \bar\delta\left(Z^*\!\cA\right)
	 =\left(\del_{\bar\beta}\cA_{\bar\alpha}\frac{\del \bar z^{\bar\beta}}{\del \bar t^{\bar\imath}}
	 \frac{\del\bar z^{\bar\alpha}}{\del\bar\sigma} 
	 + \cA_{\bar\alpha}\frac{\del^2\bar z^{\bar\alpha}}{\del\bar t^{\bar\imath}\,\del\bar\sigma}\right)
	 \rd\bar t^{\bar\imath}\wedge\rd\bar\sigma\, .
\label{dBofA}
\ee
The most important property here is that, because the map $Z$ varies holomorphically, {\it only antiholomorphic derivatives of the connection arise}. Again integrating by parts we find that
\be
\begin{aligned}
	\overline{D}_{\!\mathscr{B}}\rU(\sigma_1,\sigma_0;t) 
	&\equiv\bar\delta\rU(\sigma_1,\sigma_0;t) + \cA(z(\sigma_1,t))\rU(\sigma_1,\sigma_0;t) 
	- \rU(\sigma_1,\sigma_0;t)\,\cA(z(\sigma_0,t))\\
	 &= -\int_\Sigma \omega_{1,0}(\sigma)\wedge\rd\bar\sigma\wedge\rd\bar t^{\bar\imath}\	
	  \rU(\sigma_1,\sigma;t)\,\cF_{\bar\alpha\bar\beta}\frac{\del\bar z^{\bar\alpha}}{\del\bar \sigma}
	  \frac{\del\bar z^{\bar\beta}}{\del\bar t^{\bar\imath}}\,\rU(\sigma,\sigma_0;t)\\
	  &=-\int_C \omega_{1,0}(z)\wedge\rd\bar z^{\bar\alpha}\wedge\bar\delta\bar z^{\bar\beta}\,
	  \rU(z_1,z)\cF_{\bar\alpha\bar\beta}(z)\,\rU(z,z_0)
 \end{aligned}
\label{Uvary}
\ee
where $\overline{D}_{\!\mathscr{B}}$ is the natural covariant $\delbar$-operator acting on sections of the sheaf $\mathscr{E}$ over $\mathscr{B}$ induced by\footnote{More accurately, $\mathscr{E} = \pi_*{\rm ev}^*_\sigma E$.} $E$. 

Equation~\eqref{Uvary} tells us how the holomorphic frame $\rU(z_1,z_0)$ on an irreducible curve varies as we move around in the moduli space. For the Wilson Loop~\eqref{hWL} on a nodal curve we now readily find
\be
\begin{aligned}
	\bar\delta\rW_{\!R}[C(t)]  
	&= \sum_{i=1}^n{\rm Tr}_R\left(\Big(\overline{D}_{\!\mathscr{B}}\rU(z_{i+1},z_i)\Big)
	\rU(z_i,z_{i-1})\cdots\rU(z_{i+2},z_{i+1})\right)\\
	&=-\int_{C(t)} \omega(z)\wedge\rd \bar{z}^{\bar\alpha}\wedge \bar\delta\bar z^{\bar\beta}\ 
	{\rm Tr}_{R}\Big(\cF_{\bar\alpha\bar\beta}(z,\bar z)\,{\rm Hol}_{z}[C(t)]\Big)
\end{aligned}
\label{Wvary2}
\ee
where the integral is taken over the complete nodal curve $C(t)=Z(\Sigma)$ and we emphasise that this curve depends on the moduli of the map. As usual, $\omega(z)$ is a meromorphic differential on $C(t)$ with only simple poles at the nodes. This completes the proof.


\section{Holomorphic Loop Equations}
\label{sec:loopeq}

In this section we derive an analogue of the loop equations of Makeenko and Migdal~\cite{Makeenko:1979pb} for our complex Wilson Loops. To do this, instead of considering the complexified Wilson Loop itself, we must study its average $\big\la \rW_{\!R}[C]\big\ra$ over the space of gauge-inequivalent connection (0,1)-forms on $E$, weighted by the exponential of an appropriate action.

In section~\ref{sec:hlink} we choose the action to be the U$(N)$ holomorphic Chern-Simons functional
\be
	S_{\rm hCS}[\cA] = \frac{1}{\rm g^2}\int\limits_{\CP^{3|4}} \rD^{3|4}z\wedge
	{\rm tr}\left(\frac{1}{2}\cA\wedge\delbar \cA + \frac{1}{3}\cA\wedge\cA\wedge\cA\right)\, ,
\label{hCS}
\ee
that may be interpreted as the open string field theory of the perturbative B-model~\cite{Witten:1992fb,Witten:2003nn}. In real Chern-Simons theory on a real 3-manifold $M$, the correlation function of a Wilson Loop gives a knot invariant. Since these invariants depend only on the (regular) isotopy class of the knot, one expects that, unlike $\rW_{\!R}[\gamma]$ itself, $\big\la \rW_{\!R}[\gamma]\big\ra_{\rm CS}$ should remain invariant as one smoothly deforms $\gamma$. This is essentially true, except that there are important corrections when the deformation causes $\gamma$ to pass through a self-intersection so that the isotopy class of the knot jumps. The loop equations determine the behaviour of $\big\la\rW_{\!R}[\gamma]\big\ra_{\rm CS}$ under such a jump and it was shown in~\cite{CottaRamusino:1989rf} that, for Wilson Loops in the fundamental representation of U($N$), they amount to the skein relations\be
	s^{\frac{1}{2}}P_+(r,s) - s^{-\frac{1}{2}}P_-(r,s) = (r^{\frac{1}{2}}-r^{-\frac{1}{2}})P_0(r,s)
\label{HOMFLYskein}
\ee
from which the HOMFLY polynomial $P_\gamma(r,s)$ can be recursively constructed\footnote{The variables of the HOMFLY polynomial are related to $N$ and the level $k$ of the Chern-Simons theory by and $s=\e^\lambda$ and $r=\e^{\lambda/N}$, where $\lambda\equiv 2\pi\im N/(k+N)$  may be interpreted as the 't Hooft coupling.  In~\cite{CottaRamusino:1989rf}, the skein relations were obtained to lowest order in $\lambda$. The skein relation only determines the HOMFLY polynomial for Wilson Loops in the fundamental.}.

We expect that correlation functions of our Wilson Loops in the holomorphic Chern-Simons QFT are likewise associated with {\it holomorphic linking invariants}. Indeed, \cite{Khesin:2000ng,Frenkel:2005qk} show that for two non-intersecting curves  $C_1,\,C_2\subset\bC^3$ of genera $g_i\geq1$, the expectation value $\big\la\rW[C_1]\,\rW[C_2]\big\ra$ in Abelian holomorphic Chern-Simons theory on $\bC^3$ involves\footnote{We will see later that, at least for the $\cN=4$ theory, `holomorphic self-linking' is well-defined even without a choice of framing.} the {\it holomorphic linking invariant}
\be
	L(C_1,\theta_1;C_2;\theta_2) 
	= \frac{1}{4\pi}\int\limits_{C_1\times C_2} 
	\epsilon_{\bar\imath\bar\jmath\bar k}\frac{(\bar z-\bar w)^{\bar\imath}}{|z-w|^6}\,
	\rd \bar z^{\bar\jmath}\wedge\rd\bar w^{\bar k}\wedge \theta_1\wedge\theta_2
\label{hGauss}
\ee
that generalizes the Gauss linking number
\be
	L(\gamma_i,\gamma_j)
	 = \frac{1}{4\pi}\int\limits_{\gamma_i\times\gamma_j}\epsilon_{ijk}\frac{(x-y)^i}{|x-y|^3}\,\rd x^j\wedge\rd y^k
\label{Gauss}
\ee
in for real Abelian Wilson Loops in $\bR^3$. Observe that~\eqref{hGauss} depends on a choice of holomorphic 1-form $\theta_i$ on each curve. For our nodal curve, the corresponding form is meromorphic and depends on the location of the nodes. In section~\ref{sec:hlink} we shall find that $\bar\delta\big\la\rW_{\!R}[C]\big\ra_{\rm hCS}$ generically vanishes, with important corrections when the type of the curve jumps. In the planar limit, the loop equations may be reduced to the BCFW relations~\cite{Britto:2005fq} that recursively construct the classical S-matrix. In this sense, the BCFW relations are simply the U$(\infty)$ skein relations for the holomorphic link defined by $C$.

\medskip

As mentioned in the introduction, holomorphic Chern-Simons theory on $\CP^{3|4}$ corresponds only to the anti self-dual sector of $\cN=4$ super Yang-Mills theory on space-time. In section~\ref{sec:twistoraction} we replace~\eqref{hCS} by the action~\cite{Boels:2006ir}
\be
	S_{\cN=4}[\cA] = 
	S_{\rm hCS}[\cA] + \int_\Gamma {\rm d}^{4|8}x\,  \ln\det\!\left.(\delbar+\cA)\right|_{\rm X}
\ee
that, as our notation indicates, describes $\cN=4$ super Yang-Mills theory on twistor space. We shall see that the new term is itself eminently compatible with the Wilson Loop. The loop equations in this theory involve a new term that leads to the all-loop extension of the BCFW relations found by Arkani-Hamed {\it et al.} in~\cite{Arkani-Hamed:2010kv}.


\subsection{Holomorphic linking}
\label{sec:hlink}

In this section, we choose the action to be the holomorphic Chern-Simons functional~\eqref{hCS}, where
$\rD^{3|4}z$ is the canonical holomorphic section of the Berezinian of $\CP^{3|4}$, given explicitly by 
\be
	\rD^{3|4}z \equiv
	\frac{1}{4!}\epsilon_{\alpha\beta\gamma\delta}z^\alpha\rd z^\beta\wedge\rd z^\gamma\wedge\rd z^\delta\ 
	\frac{1}{4!}\epsilon_{abcd}\rd\psi^a\,\rd\psi^b\,\rd\psi^c\, \rd\psi^d
\ee
in terms of homogeneous coordinates $(z^\alpha,\psi^a)\in\bC^{4|4}$. This action is defined on the space $\mathscr{A}^{0,1}$ of partial connections $\bar{D}$ on a $C^\infty$ vector bundle $E\to\CP^{3|4}$, given at least locally by $\bar{D} =\delbar+\cA$ in terms of a background partial connection $\delbar$. We take this to obey $\delbar^2=0$ and so define a background complex structure on $E$. Thus, $\cA(z,\bar z,\psi)$ is a superfield  with component expansion
\be
	\cA(z,\bar z,\psi) = a(z,\bar z)+\psi^a\Gamma_a(z) + \frac{1}{2!}\psi^a\psi^b\,\Phi_{ab}(z,\bar z) 
	+ \cdots + \frac{1}{4!}\epsilon_{abcd}\psi^a\psi^b\psi^c\psi^d\,g(z,\bar z)\,,
\label{superA}
\ee
where the coefficient of $(\psi)^r$ is a (0,1)-form on $\CP^3$, valued in smooth sections of ${\rm End}(E)\otimes\cO_{\CP^3}(-r)$. 

We choose the gauge group to be U($N$) and consider only Wilson Loops in the fundamental representation. The Yang-Mills (or open string) coupling ${\rm g}^2$  is then related to the Chern-Simons level $k$ by ${\rm g}^2= 2\pi/(k+N)$. In anticipation of taking the planar limit, we write ${\rm g}^2 = \lambda/N$ in terms of the 't Hooft coupling $\lambda$ and include a factor of $1/N$ in definition of the Wilson Loop so that $\rW[C]=1$ if the holonomy is trivial.

\medskip

Inserting~\eqref{hWLvary} into the holomorphic Chern-Simons path integral, one finds
\be
\begin{aligned}
	\bar\delta\big\la \rW[C(t)]\big\ra 
	&= -\frac{1}{N}\int\cD\!\cA\,
	\left[\int_{C(t)} \omega(z)\wedge\rd \bar{z}^{\bar\alpha}\wedge \bar\delta\bar z^{\bar\beta}\ 
	{\rm tr}\Big(\cF_{\bar\alpha\bar\beta}(z)\,{\rm Hol}_{z}[C(t)]\Big)\right]\e^{-S_{\rm hCS}[\cA]}\\
	&= \frac{\lambda}{N^2} \int \cD\!\cA\,
	\left[\int_{C(t)}\omega(z)\wedge  {\rm tr} \left({\rm Hol}_z[C(t)]\
	\frac{\delta}{\delta\cA(z)}\,\e^{- S_{\rm hCS}[\cA]}\right)\right]\,,
\end{aligned}
\label{SchwDysonhCS}
\ee
since the variation of the holomorphic Chern-Simons functional is $ \cF^{0,2}\times N/\lambda$. To obtain an interesting equation for $\bar\delta\big\la\rW[C(t)]\big\ra$, as in~\cite{Makeenko:1979pb} we integrate by parts in the path integral, bringing the variation of the connection to act on the holonomy. Essentially the same argument as in section~\ref{sec:holfamily} shows that under a variation of the connection at some point $z'\in C$, (displaying colour indices)
\be
\begin{aligned}
	0&=\left[\left.\frac{\delta}{\delta\cA(z)}\right.^i_{\ j}
	\left(\int_C\omega_{1,0}(z')\wedge \rU(z_1,z')
	\left(\delbar+\cA\right)_{C}\rU(z',z_0)\right)^k_{\ l}\right]\\
	&=-\left[\left.\frac{\delta}{\delta\cA(z)}\right.^i_{\ j}\!\rU(z_1,z_0)^k_{\ l}\right]
	+ \int_C\omega_{1,0}(z')\wedge \rU(z_1,z')^k_{\ j}\,\bar\delta^{3|4}(z, z')\,\rU(z',z_0)^i_{\ l}
\label{UvaryA}
\end{aligned}
\ee
where we have used the fact that, for a U$(N)$ gauge group,
\be
	\left(\delta/\delta\cA(z)\right)^i_{\ j}\cA(z')^m_{\ \,n} = \bar\delta^{3|4}(z,z')\,\delta^i_{\ n} \delta^m_{\ \,j}
\ee
with $\bar\delta^{3|4}(z,z')$ a Dirac current concentrated on the diagonal in $\CP^{3|4}_z\times \CP^{3|4}_{z'}$, {\it i.e.} $\bar\delta^{3|4}(z,z')$ is a distribution-valued (0,3)-form on $\CP^{3|4}_z\times \CP^{3|4}_{z'}$ such that for any $\alpha\in\Omega^{0,p}(\CP^{3|4})$ we have
\be
	\int_{\CP^{3|4}} \rD^{3|4}z'\wedge\bar\delta^{3|4}(z,z')\wedge \alpha(z') = \alpha(z)\, .
\ee
Explicitly, $\bar\delta^{3|4}(z,z')$ may be represented by the integral~\cite{Mason:2009qx,Mason:2009sa}
\be
	\bar\delta^{3|4}(z,z')= \int\frac{\rd u}{u}\wedge \bar\delta^{4|4}(z-uz') 
	= \int\frac{\rd u}{u} \bigwedge_{\alpha=1}^4\delbar\frac{1}{z^\alpha-u{z'}^\alpha}\ \prod_{a=1}^4 (\psi^a-u{\psi'}^a)
\label{explicitdelta34}
\ee
that forces $z^\alpha \propto {z'}^\alpha$ so that the two point must coincide projectively.

\FIGURE[t]{
	\includegraphics[width=85mm]{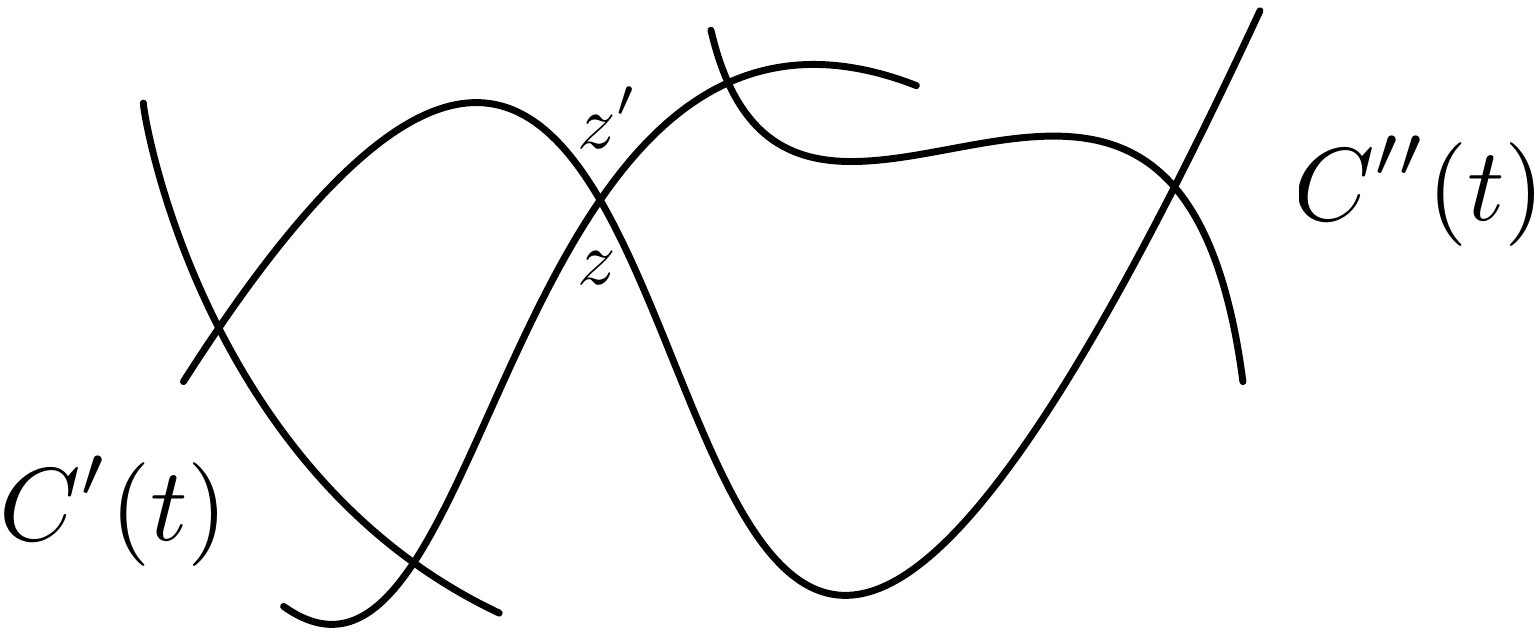}
	\caption{The expectation value $\big\la\rW[C(t)]\big\ra$ in holomorphic Chern-Simons theory varies holomorphically 
	except where the curve develops a new node.}
\label{fig:selfintersect}
}

We now apply equation~\eqref{UvaryA} to the variation holonomy based, say, at a point $z\in C_1$. One finds
\be
\begin{aligned}
	&{\rm tr}\left( \frac{\delta}{\delta\cA(z)}{\rm Hol}_z[C(t)]\right) 
	= \left.
	\frac{\delta}{\delta\cA(z)}\right.^i_{\ j}\Big[\rU(z,z_n)^j_{\ k}\rU(z_n,z_{n-1})^k_{\ l}\cdots\rU(z_1,z)^m_{\ \, i}\Big]\\
	&=\sum_i\int_{C_i(t)}\hspace{-0.2cm}\omega_{i+1,i}(z')\wedge\bar\delta^{3|4}(z,z')\ 
	{\rm tr}\Big[\rU(z,z_n)\cdots\rU(z_{i+1},z')\Big]\, {\rm tr}\Big[\rU(z',z_i)\cdots \rU(z_1,z)\Big]\ .
\end{aligned}
\label{HolvaryA}
\ee
The factor of $\bar\delta^{3|4}(z,z')$ ensures that this term has support only at points $t$ in the moduli space where the curve $C(t)$ degenerates so that the component  $C_1(t)$ either self-intersects or else intersects some other component. Similar contributions arise from summing over the possible locations of the base-point $z$.
In other words, thinking of our original curve as an elliptic curve, $\bar\delta \big\la\rW[C(t)]\big\ra$ vanishes everywhere except on boundary components of the moduli space where the holomorphic map degenerates so that two points on the source curve are mapped to the same image. 

Combining all the terms, the holomorphic Chern-Simons loop equations~\eqref{SchwDysonhCS} become
\be
	\bar\delta\big\la\rW[C(t)]\big\ra 
	=\ -\lambda \hspace{-0.4cm} \int\limits_{C(t)\times C(t)} 
	\omega(z)\wedge\omega(z')\wedge\bar\delta^{3|4}(z,z')\ 
	\big\la\rW[C^\prime(t)]\,\rW[C^{\prime\prime}(t)]\big\ra\,,
\label{loopeqnonplanar}
\ee
for normalised Wilson Loops in the fundamental of a U$(N)$ gauge group, where $C'(t)$ and $C''(t)$ are the two curves obtained by ungluing $C(t)$ at its new node $z=z'$ (see figure~\ref{fig:selfintersect}). In the planar limit, the correlation function of a product of Wilson Loops factorizes into the product of correlation functions, so~\eqref{loopeqnonplanar} simplifies to
\be
	\bar\delta\big\la\rW[C(t)]\big\ra 
	=\ -\,\lambda\hspace{-0.4cm}\int\limits_{C(t)\times C(t)}\omega(z)\wedge\omega(z')\wedge\bar\delta^{3|4}(z,z')\ 
	\big\la\rW[C^\prime(t)]\big\ra\,\big\la\rW[C^{\prime\prime}(t)]\big\ra
\label{MigMakhCS}
\ee
where $\lambda$ is the 't Hooft coupling.  Thus, in the planar limit, the behaviour at a self-intersection is determined by the product of the expectation values of the Wilson Loops around the two lower degree curves. This may be viewed as an analogue for holomorphic linking of the skein relations for real knot invariants. In section~\ref{sec:amplitudes} we shall see that the classical BCFW recursion relations are simply a special case of~\eqref{MigMakhCS}.


\subsection{Loop equations for the twistor representation of $\cN=4$ SYM}
\label{sec:twistoraction}

We now consider the expectation value of the complex Wilson Loop in the twistor action
\be
	S_{\cN=4}[\cA] = 
	S_{\rm hCS}[\cA] + \int_\Gamma {\rm d}^{4|8}x\,  \log\det\!\left.(\delbar+\cA)\right|_{\rm X}\,,
\label{twistoraction}
\ee
where X is a line in twistor space and $\Gamma$ is a middle-dimensional contour in the space ${\rm Gr}_2(\bC^4)$ of these lines\footnote{The fermionic integral is done algebraically, as always.}, interpreted as a choice of real slice of complexified, conformally compactified space-time. The proof that~\eqref{twistoraction} is an action for $\cN=4$ SYM on twistor space is given in~\cite{Boels:2006ir}, to which the reader is referred for further details. (Readers familiar with the MHV formalism may note that the new term in~\eqref{twistoraction} provides an infinite series of MHV vertices, while the 3-point $\MHVbar$ vertex in the Chern-Simons theory vanishes in an axial gauge. From the point of view of twistor string theory~\cite{Witten:2003nn}, the determinant may be understood as the partition function of a chiral free fermion CFT on X, coupled to the pullback of the twistor space gauge field.)

From the point of view of the present article, it is possible to motivate this action by observing that Wilson Loops on complex curves are again the natural observables of the QFT based on~\eqref{twistoraction}. The reason this is so is because of the close connection between determinants and holonomies. More precisely, just as
\be
	\delta \log\det M = \delta\, {\rm tr}\log M = {\rm tr}(M^{-1}\delta M)
\ee
for a finite dimensional matrix $M$, Quillen shows~\cite{QuillenLine} that, for an abstract curve $\Sigma$, under a variation of the connection, the logarithm of a section of the determinant line bundle ${\rm Det}\to\mathscr{A}^{0,1}_\Sigma$ varies as\footnote{In writing this formula, we assume that $\delta$ is a flat connection on ${\rm Det}$ in a trivialisation where the local connection 1-form vanishes. That Det should admit such a flat connection is necessary if we wish to treat $\det(\delbar+\cA)_\Sigma$ as a (holomorphic) function on $\mathscr{A}_\Sigma^{0,1}$ (the space of connection (0,1)-forms on $\Sigma$). However~\cite{QuillenLine}, the construction of this flat connection requires a choice of background connection.}
\be
	\delta\,\log\det\left.\!\!(\delbar+\cA)\right|_\Sigma
	= \int_\Sigma {\rm tr} \left(J_\cA\wedge \delta\cA\right)\,,
\label{Quillenvary}
\ee
where 
\be
	J_{\cA}(\sigma) \equiv \lim\limits_{\sigma'\to \sigma}\Big(G_\cA(\sigma',\sigma) - G_0(\sigma',\sigma)\Big)
\label{Jdef}
\ee
is the limit\footnote{Below, this limit will be revealed as nothing but the forward limit of an amplitude!} on the diagonal in $\Sigma\times\Sigma$ of difference between the Green's function for $\delbar+\cA|_\Sigma$ and the Green's function for background connection $\delbar_\Sigma$\,.   These Green's functions are non-local operators on the Riemann surface; the fact that we take their limit on the diagonal can be understood as part of the trace. For a single such Green's function, the limit on the diagonal is necessarily singular, but the singularity cancels in the difference~\eqref{Jdef}. The dependence on a background connection reflects the fact that to treat sections of the determinant line bundle as functions we must first pick a trivialisation, amounting to a choice of background connection (0,1)-form on the bundle over $\Sigma$. In our case, we will of course choose this background $\delbar$-operator to be the one induced by our choice of base-point $\delbar\in\mathscr{A}^{0,1}_\Sigma$ in writing the holomorphic Chern-Simons action.

When $\Sigma$ is a Riemann sphere, the Green's functions are
\be
	G_\cA(\sigma',\sigma) = h(\sigma')G_0(\sigma',\sigma)h^{-1}(\sigma)
	\qquad\hbox{and}\qquad
	G_0(\sigma',\sigma) = \frac{1}{2\pi\im} \frac{\rd\sigma}{\sigma'-\sigma} \,,
\ee
in terms of the holomorphic frame $h$ on $\Sigma$. Therefore Quillen's prescription reduces to
\be
	J_\cA(\sigma) = \frac{\rd \sigma}{2\pi\im}\, 
	\lim\limits_{\sigma'\to \sigma}\,\frac{\rU(\sigma',\sigma)-\rU(\sigma,\sigma)}{\sigma'-\sigma}
	= \frac{\rd \sigma}{2\pi\im}\left. \frac{\del\rU(\sigma',\sigma)}{\del \sigma'}\right|_{\sigma'=\sigma} \,.
\label{JUrelation}
\ee
For our purposes, the presence of the holomorphic derivative here is somewhat inconvenient. However, it is easily seen from the formal series~\eqref{Pexp2} that $\rU(\sigma',\sigma)$ depends only meromorphically on $\sigma'$ and is regular at $\sigma'=\sigma$. Thus we can use Cauchy's theorem to rewrite the derivative in~\eqref{JUrelation} as an integral, so
\be
\begin{aligned}
	\delta\,\log\det\!\left.(\delbar+\cA)\right|_\Sigma 
	&= \frac{1}{(2\pi\im)^2}\int\limits_{\Sigma}\rd \sigma\wedge 
	{\rm tr} \left(\oint\frac{\rd \sigma'}{(\sigma-\sigma')^2}  \rU(\sigma',\sigma)\,\delta\cA(\sigma)\right)\\
	&=\frac{1}{(2\pi\im)^3}\int\limits_{\Sigma\times S^1\times S^1}
	\frac{\rd \sigma\wedge\rd\sigma'\wedge\rd\sigma''}{(\sigma-\sigma')(\sigma'-\sigma'')(\sigma''-\sigma)}
	{\rm tr}\left(\rU(\sigma',\sigma)\,\delta\cA(\sigma)\right)
\end{aligned}
\label{logdetvar1}
\ee
where the integrals over $\sigma''$ and $\sigma'$ are performed over contours encircling the poles at $\sigma''=\sigma'$ and $\sigma'=\sigma$, respectively. (The reason for introducing the new point $\sigma''\in\Sigma$ will become clear momentarily.)

To apply this to the case that $\Sigma = Z^{-1}({\rm X})$ for X a line in twistor space, suppose
\be
	Z(\sigma'') = z_a\qquad Z(\sigma')= z_b\qquad\hbox{and}\qquad Z(\sigma) =\hat z=z_a+\sigma z_b
\ee
so that ${\rm X}={\rm Span}[z_a,z_b]$. Then
\be
	\frac{\rd \sigma\wedge\rd\sigma'\wedge\rd\sigma''}{(\sigma-\sigma')(\sigma'-\sigma'')(\sigma''-\sigma)}
	=Z^*\!\left[\omega_{a,b}(\hat z)\wedge\frac{\la a\,\rd a\ra\wedge\la b\,\rd b\ra}{\la a\,b\ra^2}\right]
\ee
where $\omega_{a,b}(\hat z)$ is meromorphic with only simple poles only $\hat{z}=z_a $ and $z_b$, and where 
$\la a\,\rd a\ra\la b\,\rd b\ra/\la a\,b\ra^2$ is the fundamental bi-differential on X with a double pole along the diagonal.
This bi-differential combines with the integral over the space of lines X, 
since
\be
	\rd^{4|8}x\wedge\frac{\la a\,\rd a\ra\wedge\la b\,\rd b\ra}{\la a\,b\ra^2}= \rD^{3|4}z_a\wedge\rD^{3|4}z_b\,,
\ee
giving a contour integral over $\CP^{3|4}_a\times\CP^{3|4}_b$. Therefore, the variation of the new term in the action is 
\be
	\delta\int\limits_{\Gamma}\rd^{4|8}x\,\log\det\!\left.(\delbar+\cA)\right|_{\rm X} 
	=\int\limits_{\Gamma\times {\rm X}\times S^1\times S^1}\hspace{-0.4cm}
	\frac{\rD^{3|4}z_a\wedge\rD^{3|4}z_b}{(2\pi\im)^3}\wedge\omega_{a,b}(\hat z)\,
	{\rm tr}\Big(\rU(z_b,\hat{z})\,\delta\cA(\hat{z})\Big)\,,
\label{logdetvar}
\ee
where the $S^1\times S^1$ contour sets $z_a\to \hat{z}$ and $z_b\to\hat z$, differentiating $\rU(z_b ,\hat z)$ in the process. The integral over all $\hat z\in {\rm X}$ is performed using the dependence on $\hat z$ in the holomorphic frame and in the variation of the connection.

Equation~\eqref{logdetvar} expresses the variation of the logarithm of the determinant in the twistor action~\eqref{twistoraction} in terms of a holomorphic frame on the line X. It dovetails beautifully with the holonomy around the curve $C$, providing a new contribution
\be
	-\frac{\lambda}{N}\hspace{-0.4cm}
	\int\limits_{\Gamma\times S^1\times S^1}\hspace{-0.4cm} \rD^{3|4}z_a\wedge\rD^{3|4}z_b\left[\,
	\int\limits_{C(t)\times{\rm X}} \omega(z)\wedge \omega_{a,b}(\hat z)\wedge\bar\delta^{3|4}(z,\hat z) 
	\,{\rm tr}\Big(\rU(z_b,\hat{z})\,{\rm Hol}_z[C(t)]\Big)\right]
\label{extraterm}
\ee
to the loop equations. The $\delta$-function $\bar\delta^{3|4}(z,\hat z)$ in this expression  means that this term only contributes at points in the moduli space where the curve $C(t) \owns z $ intersects the line ${\rm X}\owns \hat z$. The integral over $\Gamma$ then adds up these contributions for each ${\rm X}\in\Gamma$.

To interpret the product $\rU(z_b,\hat z)\,{\rm Hol}_{z}[C(t)]$, replace $C(t)$ by a curve $\widetilde{C(t)}$ where
\be
	\widetilde{C(t)}\cap{\rm X} = \{\hat z,z_b\}\in\CP^{3|4}
\ee
and such that as $z_b\to \hat{z}$, $\widetilde {C(t)}\to C(t)$. As a nodal curve, $\widetilde{C(t)}$ has one more component than $C(t)$, with one component becoming double covered in the limit  (see figure~\ref{fig:fwdlim}). Now, because the contour integral forces $z_b=\hat{z}$, we can replace the holonomy around $C(t)$, based at $z=\hat{z}=C(t)\cap{\rm X}$, by a product of holomorphic frames that transport us around  $\widetilde{C(t)}$ from $z_b$ to $\hat{z}$. The remaining factor of $\rU(z_b,\hat{z})$ transports this frame back to $z_b$ along X. The resulting trace can be interpreted as Wilson Loop around $\widetilde {C(t)}\cup {\rm X}$, so that
\be
	\frac{1}{N}\,{\rm tr} \Big(\rU(z_b,\hat{z}){\rm Hol}_z[C(t)]\Big)
	=\rW[\widetilde{C(t)}\cup{\rm X}] \,,
\ee
{\it provided we are inside the contour integral over the location of} $z_b\subset {\rm X}$.

\medskip
\FIGURE[t]{
	\includegraphics[height=45mm]{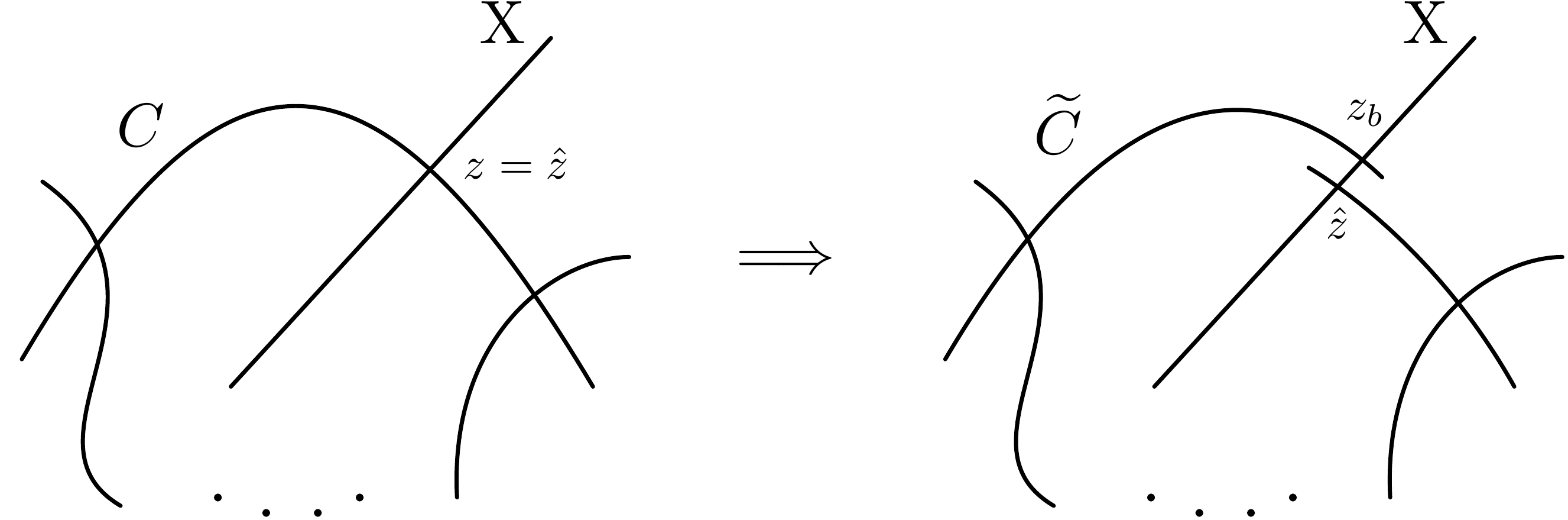}	
	\caption{\small The new term in the loop equations for $\cN=4$ SYM can be understood as a holomorphic Wilson 
	Loop around a new curve $\widetilde C\cup{\rm X}$, where $\widetilde C$ reduces to $C$ when $z_b\to \hat z$
	along X. Note that the nodal curve $\widetilde C$ has one more component than $C$. 
	In the scattering amplitude context,  $\la \rW[\widetilde C\cup{\rm X}]\ra$ may be interpreted as a forward limit of an 
	$n+2$ particle amplitude.}
\label{fig:fwdlim}
}

Finally then, we see that the holomorphic loop equations for the twistor representation of planar $\cN=4$ SYM are
\be
\begin{aligned}
	\bar\delta\big\la \rW[C(t)]\big\ra
	&=\  -\ \lambda \hspace{-0.2cm}
	\int\limits_{C(t)\times C(t)}\hspace{-0.2cm}\omega(z)\wedge\omega(z')\wedge\bar\delta^{3|4}(z, z')\,
	\big\la\rW[C^\prime(t)]\big\ra\,\big\la\rW[C^{\prime\prime}(t)]\big\ra\\
	&\hspace{-0.4cm}-\ \lambda\hspace{-0.2cm}
	\int\limits_{\Gamma\times S^1\times S^1}\hspace{-0.4cm} \rD^{3|4}z_a\wedge\rD^{3|4}z_b\left[\,
	\int\limits_{C(t)\times{\rm X}} \hspace{-0.2cm}\omega(z)\wedge \omega_{a,b}(\hat z)\wedge\bar\delta^{3|4}(z,\hat z) \,
	\big\la\rW[\widetilde{C(t)}\cup{\rm X}]\big\ra\right]
\end{aligned}
\label{MigMakN=4}
\ee
and, as in the original equations of Migdal \& Makeenko~\cite{Makeenko:1979pb}, are entirely expressed in terms of expectation values of our (complexified) Wilson Loops. In the following section we show that for a particular one-parameter family of curves $C(t)$, this holomorphic loop equation reduces to the all-loop BCFW recursion relation found in~\cite{Arkani-Hamed:2010kv}, but we emphasise that~\eqref{MigMakN=4} is valid for a far more general class of curves.


\section{BCFW Recursion from the Loop Equations}
\label{sec:amplitudes}

We now apply the holomorphic loop equation~\eqref{MigMakN=4} to the case that each component of $C$ is a line, as in figure~\ref{fig:knot}.  $C$ can be completely specified by giving the location of its $n$ nodes, and we shall often write $C=(12)\cup(23)\cup\cdots\cup(n1)$ for the curve (where $(i\,i\!+\!1) \equiv {\rm Span}[z_i,z_{i+1}]$) and $\rW[1,2,\ldots,n]$ for the corresponding Wilson Loop (in the fundamental representation). We wish to show that the loop equations for $\big\la \rW[1,2,\ldots,n]\big\ra$ directly reduce to the all-loop BCFW recursion relations of Arkani-Hamed {\it et al.}~\cite{Arkani-Hamed:2010kv}, although it is important that our derivation of these relation is completely independent of any notions of scattering theory. 

Consider the holomorphic family of Wilson Loops determined by translating the node $z_n$ along the line $(n\!-\!1,n)$. That is, we define
\be
	\widehat{z}_n(t) \equiv z_n+tz_{n-1}
\label{Zhatdef}
\ee
and consider the 1-parameter family of holomorphic curves
\be
	C(t) \equiv (12)\cup(23)\cup\cdots\cup (n\!-\!1\,\hat{n}(t))\cup(\hat n(t)\,1)
\ee
as shown in figure~\ref{fig:BCFWdef}. Notice that although $t$ can be thought of as a local holomorphic coordinate on the line $(n\!-\!1\,n)$, it is really a coordinate on the moduli space of our family of curves, {\it i.e.} a coordinate on a $\CP^1\subset\mathscr{B}$. As $t$ varies, the line $(\hat{n}(t),1)$ sweeps out the plane $(n\!-\!1,n,1)$ with all other components of $C(t)$ remaining fixed. Lines and planes necessarily intersect\footnote{Lines and planes do not necessarily intersect in the superspace $\CP^{3|4}$, because they might `miss' in the fermionic directions. The fermionic $\delta$-functions in the $\bar\delta^{3|4}(z,z')$ ensure that the right hand side has support only when the curves do intersect in the superspace. See also the discussion in~\cite{Bullimore:2010pj}.} in $\CP^3$ so for every $j=3,\ldots,n\!-\!1$ there exists a $t_j$ for which $(\hat{n}(t_j), 1)$ intersects the component $C_j$. Let us call this intersection point $I_j$ and write $n_j$ for the point $\hat{z}_n(t_j)$. Clearly, we have
\be
	I_j = (n\!-\!1,n,1)\cap(j\!-\!1,j)
	\qquad\hbox{and}\qquad
	\hat n_j = (n\!-\!1,n)\cap (j\!-\!1,j,1)\, ,
\label{selfintersectionpoint}
\ee
essentially by definition. (Note that we do not get any new intersections when $j=n,1,2$.) 

\FIGURE[t]{
	\includegraphics[width=84mm]{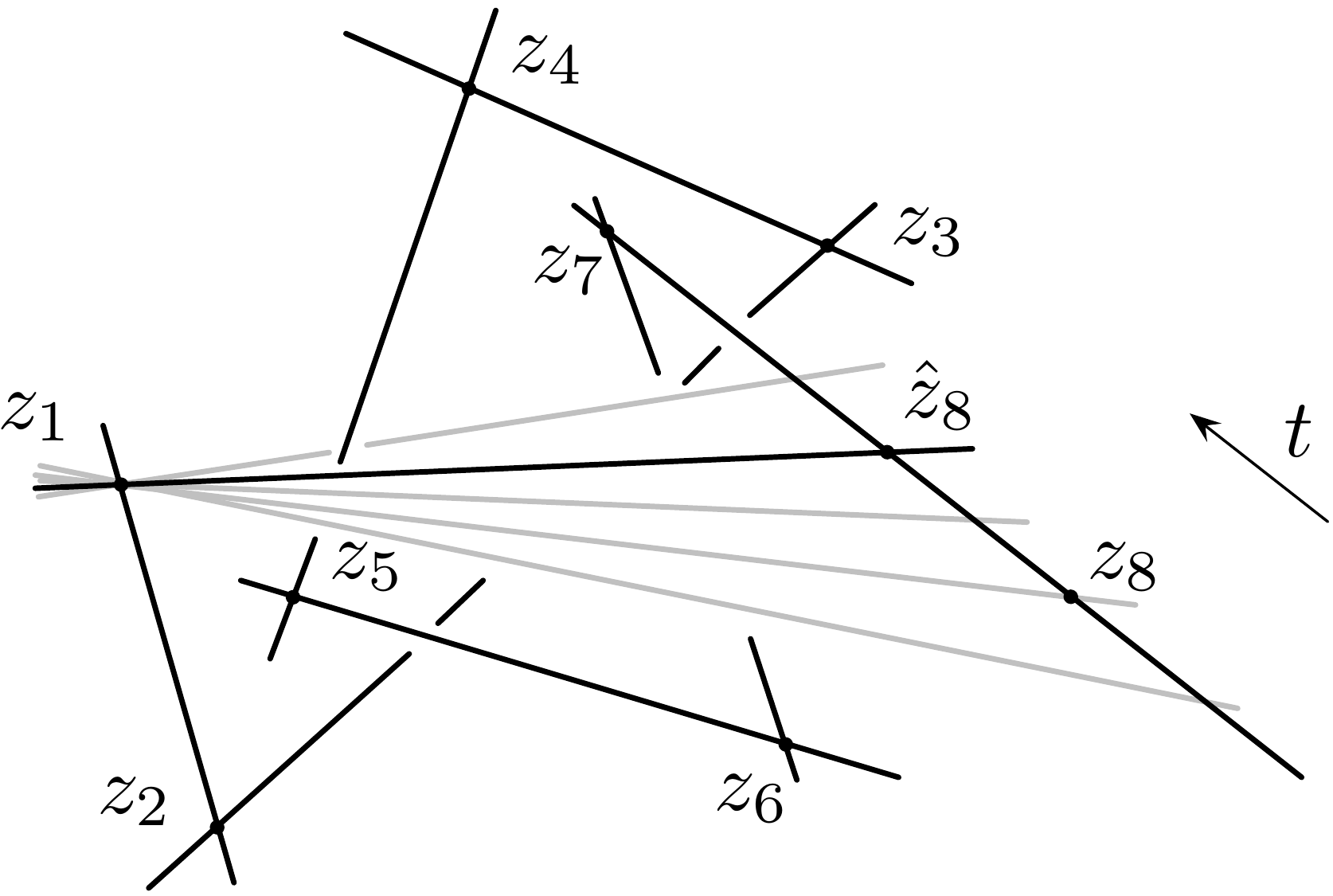}
	\caption{The family of nodal curves $C(t)$ corresponding to a BCFW deformation of a planar amplitude in 
	$\cN=4$ SYM, shown for $n=8$. As we vary over the moduli space, the line $(1,\hat n(t))$ sweeps out a pencil of 
	lines in the plane $(n\!-\!1,n,1)$ with focus $z_1$. This plane necessarily intersects every other component of $C(t)$.}
\label{fig:BCFWdef}
}

\medskip

We now study the loop equations for this family. Considering first the case of pure holomorphic Chern-Simons theory, \eqref{MigMakhCS} becomes
\be
\begin{aligned}
	&-\int_{\CP^1}\frac{\rd t}{t} \wedge\bar\delta\big\la \rW[C(t)]\big\ra\ =\\
	&\hspace{1cm} \lambda\ \sum_{j=3}^{n-1}\,
	\int\limits_{\CP^1\times C_1(t)\times C_j}\hspace{-0.2cm}
	\frac{\rd t}{t}\wedge\omega(z)\wedge\omega(z')\wedge\bar\delta^{3|4}(z,z')\ 
	\big\la\rW[C^\prime(t)]\big\ra\, \big\la\rW[C^{\prime\prime}(t)]\big\ra
\end{aligned}
\label{BCFWhCS}
\ee
where 
\be
\begin{aligned}
	C^\prime(t) &= (12)\cup(23)\cup\cdots \cup(j\!-\!1,I_j)\\
	C^{\prime\prime}(t) &= (I_j, j)\cup (j,j\!+\!1)\cup\cdots (\hat{n}_j, 1)
\end{aligned}
\ee
and where we have integrated~\eqref{MigMakhCS} over our moduli space $\CP^1\subset\mathscr{B}$ using the meromorphic differential $\rd t/t$. We can dispose of the left hand side immediately: integrating by parts gives 
\be
\begin{aligned}
	-\int_{\CP^1}\frac{\rd t}{t} \wedge\bar\delta\big\la \rW[C(t)]\big\ra
	&=\big\la\rW[C(0)]\big\ra -\big\la\rW[C(\infty)]\big\ra\\
	&=\big\la\rW[1,2,\ldots,n]\big\ra - \big\la\rW[1,2,\ldots,n\!-\!1]\big\ra\,,
\end{aligned}
\label{lhs}
\ee
the difference of the Wilson Loops around the original curve and the curve with $\hat{z}_n=z_{n-1}$.  Notice that the holomorphic frame $\rU(\hat{z}_n,z_{n-1})\to1$ as $\hat{z}_n\to z_{n-1}$ so that while the line $(n\!-\!1,n)$ does not simply `disappear' from the picture, neither does it contribute to the holonomy.

The right hand side of~\eqref{BCFWhCS} is almost as straightforward to compute, because the integrals over both the $\CP^1$ moduli space and the two copies of $C(t)$ are completely fixed by the $\bar\delta$-function $\bar\delta^{3|4}(z,z')$. Since $z\in (\hat n(t),1)$ and $z' \in (j\!-\!1,j)$ we can parametrize these integrals by setting
\be
	z = \hat{z}_n(t) + sz_1 = z_n+tz_{n-1}+sz_1\qquad\hbox{and}\qquad z' = z_{j-1}+rz_j
\label{localparam}
\ee
in terms of local coordinates $r$ and $s$ on $C_1(t)$ and $C_j$, respectively. With these coodinates, the meromorphic forms $\omega(z)$ and $\omega(z')$ become simply
\be
	\omega(z) = \frac{\rd s}{s}\qquad\hbox{and}\qquad\omega(z')=\frac{\rd r}{r} 
\ee
which indeed have simple poles at the nodes of $C_1(t)$ and $C_j$ as required. Performing the integral is then merely a matter of using the explicit form~\eqref{explicitdelta34} of $\bar\delta^{3|4}(z,z')$ and computing a Jacobian. In fact, this integral is
\be
	[n\!-\!1,n,1,j\!-\!1,j] \equiv \int\frac{\rd r}{r}\frac{\rd s}{s}\frac{\rd t}{t}\,\frac{\rd u}{u}\ 
	\bar\delta^{4|4}\!\left(z_n+tz_{n-1}+sz_1+uz_{j-1}+rz_j\right)
\label{Rinv}
\ee
which was shown in~\cite{Mason:2009qx} to be just the basic dual superconformal invariant $R_{1;jn}$ of~\cite{Drummond:2008vq}. Including the product of the two Wilson Loops on the smaller curves we find that the loop equations in pure holomorphic Chern-Simons theory reduce to
\be
\begin{aligned}
	\big\la\rW[1,\ldots,n]\big\ra &= \big\la\rW[1,\ldots,n\!-\!1]\big\ra \\
	&\hspace{-1cm}+\lambda\ \sum_{j=3}^{n-1}\ [n\!-\!1,n,1,j\!-\!1,j]\,\big\la\rW[1,\ldots,j\!-\!1,I_j]\big\ra\,
	\big\la\rW[I_j,j,\ldots,n\!-\!1,\hat{n}_j]\big\ra\, ,
\end{aligned}
\label{treeBCFW}
\ee
where $I_j$ and $\hat{n}_j$ were given in~\eqref{selfintersectionpoint}. Re-interpreting the twistor space as {\it momentum} twistor space, this is just the tree-level BCFW recursion relation~\cite{Britto:2005fq} in the momentum twistor form given in~\cite{Arkani-Hamed:2010kv}, summed over all MHV degrees. Expanding~\eqref{treeBCFW} in powers of the Grassmann coordinates $\psi$ at each vertex gives the BCFW relations for specific N$^k$MHV partial amplitudes.

\medskip

Turning now to the full $\cN=4$ theory, from~\eqref{MigMakN=4} we have an additional contribution
\be
	\lambda\hspace{-0.2cm}\int\limits_{\Gamma\times S^1\times S^1}\hspace{-0.4cm} \rD^{3|4}z_a\wedge\rD^{3|4}z_b
	\left[\,\int\limits_{\CP^1\times C(t)\times{\rm X}} \hspace{-0.2cm}\frac{\rd t}{t}\wedge
	\omega(z)\wedge \omega_{a,b}(\hat z)\wedge\bar\delta^{3|4}(z,\hat z) \,
	\big\la\rW[\widetilde{C(t)}\cup{\rm X}]\big\ra\right]\,,
\ee
where
\be
	\widetilde{C(t)}\cup{\rm X} = (12)\cup\cdots(n\!-\!1,\hat{n}(t))\cup(\hat{z},b)\cup(b,1)
\ee	
is our $n+2$ component curve. Once again, the integrals inside the square brackets are completely fixed by the $\bar\delta^{3|4}(z,\hat{z})$. The intersection point $\hat {z}$ and the point $\hat n(t)$ are fixed to be
\be
	\hat{z} = (a,b)\cap (n\!-\!1,n,1)\qquad\hbox{and}\qquad \hat{n}(t) = (n\!-\!1,n)\cap(a,b,1)
\ee
and the Jacobian from the $t$, $z$ and $\hat z$ integrals gives the $R$-invariant $[n\!-\!1,n,1,a,b]$. The $\cN=4$ loop equations thus reduce to
\be
\begin{aligned}
	\big\la\rW[1,\ldots,n]\big\ra &= \big\la\rW[1,\ldots,n\!-\!1]\big\ra \\
	&\hspace{-1cm} +\ \lambda\ \sum_{j=3}^{n-1}\ [n\!-\!1,n,1,j\!-\!1,j]\,\big\la\rW[1,\ldots,j\!-\!1,I_j]\big\ra\,
	\big\la\rW[I_j,j,\ldots,n\!-\!1,\hat{n}_j]\big\ra\\
	&\hspace{-1cm}+\ \lambda \hspace{-0.2cm}
	\int\limits_{\Gamma\times S^1\times S^1}\hspace{-0.4cm} \rD^{3|4}z_a\wedge\rD^{3|4}z_b\ [n\!-\!1,n,1,a,b\,]\,
	\big\la\rW[1,\ldots,n\!-\!1,\hat n_{ab},\hat{z},z_b]\big\ra\, ,
\end{aligned}
\label{all-loopBCFW}
\ee
where we remind the reader that the $S^1\times S^1$ contour is taken to fix $z_{a,b}\to \hat z$ along X (the $R$-invariant has a simple zero in this limit, cancelling one of the factors of $\la z_a\, z_b\ra$ in the denominator of the measure). Re-interpreting the twistor space as momentum twistor space, this is the extension of the BCFW recursion for the all-loop integrand found by Arkani-Hamed {\it et al.} in~\cite{Arkani-Hamed:2010kv}.

\medskip

Let us conclude with a few remarks. Firstly, the explicit power of the 't Hooft coupling $\lambda$ is present in~\eqref{all-loopBCFW} because our normalisation of the twistor action corresponds to the normalisation $\frac{1}{4{\rm g}^2}\int {\rm tr}(F\wedge*F) = \frac{N}{4\lambda}\int{\rm tr} (F\wedge *F)$ of the space-time Yang-Mills action. In this normalisation, $n$-particle, $\ell$-loop scattering amplitudes are proportional to $\lambda^{\ell-1} N^\chi$ (with $\chi=2$ for planar diagrams). If one rescales the twistor (or space-time) connection as $\cA\to \sqrt\lambda\cA$,  the explicit $\lambda$ disappears from~\eqref{all-loopBCFW} and the $\ell$-loop planar scattering amplitude corresponds to the coefficient of $\lambda^{(n+2\ell-2)/2}$.

Secondly, when computing real knot (rather than link) invariants, it is necessary to choose a `framing' -- roughly, a thickening of the knot into a band. For Wilson Loops in real Chern-Simons theory, this may be understood perturbatively as a point-splitting procedure that regularises potential divergences as the two ends of a propagator approach each other along the knot $\gamma$. This would seem to be particularly important in the case that $\gamma$ has cusps. Above, however,  we made no attempt to choose a `holomorphic framing' of $C$. In fact, at least for the pure holomorphic Chern-Simons theory, this does not seem to be necessary. Consider for example a propagator stretched between two adjacent line components of $C$, say $(i\!-\!1,i)$ and $(i,i\!+\!1)$. We can regularise any potential divergence in this propagator by ungluing the node, replacing $(i,i\!+\!1)\to(i_\epsilon,i\!+\!1)$ where  $z_{i_\epsilon} = z_i+\epsilon z'$ for some arbitrary point $z'$. As shown in~\cite{Mason:2010yk}, the integral of an axial gauge holomorphic Chern-Simons propagator along two non-intersecting lines gives an $R$-invariant, and in this case we obtain $[*,i\!-\!1,i,i_\epsilon,i\!+\!1]$ (where $z_*$ determines the axial gauge). It is then easy to show that this is  of order $\epsilon$ as $\epsilon\to 0$, as a consequence of the $\cN=4$ supersymmetry. In other words, the short-distance singularity of the propagator is always integrable, provided one sums over the complete $\cN=4$ supermultiplet.

If the holomorphic Chern-Simons expectation value is finite (for generic $z_i$), the same cannot be said of the expectation value in the full theory. The integral over the space of lines X will diverges (at least if $\Gamma$ is a contour corresponding to a real slice of space-time), with the divergence coming from those ${\rm X}\in\Gamma$ that intersect $C$ {\it at more than one place}, or if the intersection occurs at a node, and reflecting the divergence of the corresponding scattering amplitudes in the infra-red.  Consequently, without a regularisation (or else a choice of `leading singularity' contour for $\Gamma$), the expression~\eqref{all-loopBCFW} is somewhat formal, and provides a recursion relation for the all-orders {\it integrand}, rather than the all-orders Wilson Loop  itself. It would be fascinating if it is possible to regularise these divergences using some form of holomorphic framing of $C(t)$. Curiously, the Coulomb branch regulator of~\cite{Alday:2009zm} -- interpreted in dual conformal space-time as bringing the vertices of the piece-wise null polygon into the interior of AdS$_5$ -- corresponds in twistor space to replacing the line X by a (complex) quadric. See {\it e.g.}~\cite{Hodges:2010kq,Mason:2010pg,Alday:2010jz,Drummond:2010cz} for discussions of this regulator in a twistor context.

Further, although we considered only the particular deformation $z_n\to\hat z_n(t)$ relevant for BCFW recursion, it should be clear that the loop equations can equally be applied to the more general deformation~\cite{Elvang:2008vz}
\be	
	z_i \rightarrow z_i - t c_iz_*\qquad i=1,\ldots,n\,,
\ee
of the scattering amplitude curve, that translates each vertex towards some arbitrary point $z_*\in\CP^{3|4}$. (Here, $c_i$ are arbitrary complex constants.) This one-parameter deformation includes both BCFW recursion and the MHV diagram formalism as special cases and the validity of the corresponding recursion relation for the all-loop $\cN=4$ integrand has been shown recently in~\cite{Bullimore:2010dz}. It is also clear that many other deformations of $C$ are possible and it would be interesting to know if anything useful can be learned from, say, non-linear or multi-parameter deformations.

\medskip

As a final remark, many possible extensions to this work suggested by appealing either to the analogy with real Chern-Simons and knot theory, or else to the link to scattering amplitudes in space-time. We hope that this close relationship helps to fertilise new progress, both in understanding holomorphic linking and in $\cN=4$ SYM.

\acknowledgments

We thank Freddy Cachazo, Jaume Gomis and Natalia Saulina for helpful comments. We would also particularly like to thank Lionel Mason for many useful discussions. The work of DS is supported by the Perimeter Institute for Theoretical Physics. Research at the Perimeter Institute is supported by the Government of Canada through Industry Canada and by the Province of Ontario through the Ministry of Research $\&$ Innovation. The work MB is supported by an STFC Postgraduate Studentship.

\bibliographystyle{JHEP}
\bibliography{loopeqns}

\end{document}